# Directional asymmetry of atmospheric pole-to-pole teleconnection

Yechul Shin and Jong-Seong Kug*

School of Earth and Environmental Sciences, Seoul National University, Seoul, Republic of Korea

*Corresponding author: Jong-Seong Kug (jskug@snu.ac.kr)

## Abstract

Paleoclimate proxies show that abrupt Greenland changes were followed by Antarctic responses within decades, suggesting a fast atmospheric pole-to-pole teleconnection involving shifts in the Intertropical Convergence Zone (ITCZ) and reorganization of extratropical jets. Here, we examine how the mean ITCZ position modulates this interhemispheric teleconnection. Using an aquaplanet slab-ocean model, we apply northern high-latitude forcing in climatologies with different mean ITCZ positions. The southern high-latitude response is about twice as large when the mean ITCZ is in the Southern Hemisphere. The teleconnection is therefore directionally asymmetric. The asymmetry arises in the subtropical jet response per degree of ITCZ shift, because, even with a comparable Hadley cell response, the angular momentum carried into the upper branch depends on the distance of the ITCZ from the equator, so that an ITCZ shift toward or away from the equator reinforces or offsets the jet response. Our results suggest that the southward-displaced glacial ITCZ made Arctic-to-Antarctic transmission more efficient than today, whereas the present-day ITCZ favors the reverse.

## 1. Introduction

The Arctic and Antarctic climate systems, despite lying at opposite ends of the Earth, are dynamically connected across a wide range of timescales. Paleoclimate proxy records indicate that, during the last glacial period, warm intervals in the northern high-latitude generally coincided with cold intervals in the southern high-latitude, and vice versa, on millennial timescales (e.g., Buizert et al. 2015). This asynchronous interhemispheric pattern, known as the bipolar seesaw, has been interpreted as a consequence of ocean-mediated heat redistribution between the hemispheres. In particular, changes in the Atlantic Meridional Overturning Circulation (AMOC) are thought to play a central role in shaping this pattern (Stocker and Johnsen 2003, Pedro *et al* 2018).

The ocean, however, is not the only pathway connecting the two poles. The atmosphere can connect climate signals between them on much shorter timescales. The atmospheric pole-to-pole teleconnection proceeds in two stages. First, a high-latitude thermal forcing perturbs the tropical circulation, most clearly as a shift of the Intertropical Convergence Zone (ITCZ). The extratropical thermal signal propagates equatorward through mid-latitude eddies and lower-tropospheric advection, altering tropical sea surface temperatures (SSTs) (Hwang *et al* 2021, Shin *et al* 2021, Kim *et al* 2022). The Hadley circulation then transports the surplus energy toward the energy-deficient hemisphere, thereby reducing the interhemispheric asymmetry and shifting the ITCZ meridionally (e.g., Kang et al. 2008, 2018). Second, once tropical anomalies emerge, they can influence the opposite polar region through zonally asymmetric and symmetric pathways. In the zonally asymmetric pathway, tropical SST and convection anomalies can excite Rossby wave trains that propagate into the extratropics and affect high-latitude climate in the opposite hemisphere (e.g., Lee 2012; Ding et al. 2014; England et al. 2020; Shin et al. 2024; Moon et al. 2026). In the zonal-mean pathway, the tropical circulation change modifies the atmospheric angular momentum budget and reorganizes the subtropical and eddy-driven jets of the opposite hemisphere (Lee *et al* 2011, Ceppi *et al* 2013). The associated secondary circulations can then warm or cool in the opposite polar region adiabatically (Shin and Kang 2021; hereafter SK21).

Recent paleoclimate records reveal that Greenland and Antarctic temperatures were linked on decadal timescales during abrupt climate changes of the last glacial period. Volcanic synchronization of Greenland and Antarctic ice cores enables a more precise comparison of the two records, showing that rapid Antarctic warming occurred within decades of abrupt Greenland warming, with corresponding rapid cooling following abrupt Greenland cooling (Buizert *et al* 2018, Svensson *et al* 2026). This decadal-scale synchronous response is superimposed on the millennial-scale bipolar seesaw, and the two processes together shape the evolution of polar temperatures. Paleoclimate simulations of Dansgaard–Oeschger transitions reproduce this synchronous thermal response between the Arctic and Antarctic and attribute it to a global-scale reorganization of the zonal-mean atmospheric circulation (Svensson *et al* 2026), suggesting that the fast atmospheric pole-to-pole teleconnection operated at least during the last glacial period.

The atmospheric pole-to-pole teleconnection is relevant not only to past climates but also to present and future climate changes. For example, the delayed warming of the Southern Ocean (Armour *et al* 2016, Hwang *et al* 2017, Kug *et al* 2022) and the Antarctic meltwater discharge (Bronselaer *et al* 2018, Oh *et al* 2020, Li *et al* 2023) could perturb the tropics and hence initiate the teleconnection (Kim *et al* 2022, Kang *et al* 2023, Dong *et al* 2026). Whether such a source exists is therefore not in doubt; whether the efficiency of the teleconnection depends on the background climatology is. Glacial and present-day climates differ in many respects, and one prominent difference is the mean ITCZ position, which was shifted southward during the last glacial period relative to today (Donohoe *et al* 2013, Schneider *et al* 2014). Because the ITCZ links the forced tropical response to the circulation adjustment of the opposite hemisphere, the climatological ITCZ position is a natural candidate for controlling the efficiency.

Here, we examine how the climatological ITCZ position controls the zonal-mean atmospheric pole-to-pole teleconnection. We conduct a suite of idealized aquaplanet experiments, showing that the teleconnection is directionally asymmetric— about twice as efficient when the mean ITCZ lies in the unforced hemisphere. This asymmetry arises in the sensitivity of the subtropical jet to the ITCZ displacement, set by angular momentum conservation and hence by the distance of the ITCZ from the equator. Our results suggest that the southward-displaced glacial ITCZ made the transmission of abrupt Arctic changes to Antarctica more efficient than today.

## 2. Methods

To examine how the atmospheric pole-to-pole teleconnection depends on the background climatology, we employ the Geophysical Fluid Dynamics Laboratory atmospheric general circulation model (AM2) (The GFDL global atmospheric model development team 2004) coupled to an aquaplanet slab ocean model. The model is integrated at a horizontal resolution of 2° latitude by 2.5° longitude with 24 vertical levels. The heat capacity of the slab ocean is specified as $4 \times 10^7$ J K$^{-1}$ m$^{-2}$, corresponding approximately to a 10 m mixed-layer depth. Present-day annual-mean insolation is prescribed, with no seasonal cycle and no sea ice.

We conduct a series of experiments closely following SK21, which imposed a periodic high-latitude thermal forcing at one pole and examined the periodic response at the opposite pole. In this study, we apply the experimental design across different climatologies with systematically displaced mean ITCZ positions. We impose a surface heat flux forcing (Qflux) into the slab ocean with periodicity:

$$Q_p(t, \varphi) = A_p \sin\left(\frac{2\pi}{T} t\right) H(\varphi - \varphi_0). \quad (1)$$

The Heaviside function $H(x)$ limits the Qflux to the region poleward of $\varphi_0$, which we set to 40°N. We set the maximum amplitude $A_p$ = 10 W m$^{-2}$ and the period $T$ = 10 years. The amplitude and period are chosen so that $Q_p$ produces a detectable tropical response and, in turn, a response at the opposite pole (Fig. 1).

At the same time, a time-invariant, hemispherically asymmetric Qflux ( $Q_c$ ) is superimposed to displace the mean ITCZ:

$$Q_c(\varphi) = \begin{cases} A_c \sin\left(\frac{\varphi - 40^\circ}{50^\circ}\right), & \text{for } 90^\circ > \varphi > 40^\circ, \\ A_c \sin\left(\frac{\varphi + 40^\circ}{50^\circ}\right), & \text{for } -90^\circ < \varphi < -40^\circ, \\ 0, & \text{otherwise.} \end{cases} \quad (2)$$

Here $\varphi$ is the latitude in degrees and $A_c$ sets the magnitude and sign of the imposed interhemispheric energy contrast. For $A_c > 0$, it warms poleward of 40°N and cools poleward of 40°S. This forcing is commonly used to create an interhemispheric energy contrast and shift the ITCZ (e.g., Kang et al. 2008; Chen et al. 2025). We vary $A_c$ from -7.5 to 7.5 W m$^{-2}$ at 1.5 W m$^{-2}$ intervals, producing 11 different climatologies whose mean ITCZ spans approximately from 4°S to 4°N (Fig. 1a-c). Experiments with positive $A_c$ are denoted N1.5 to N7.5 (e.g., N1.5 for $A_c$ = 1.5 W m$^{-2}$) as their climatological ITCZ is in the Northern Hemisphere (NH), whereas those with negative $A_c$ are denoted S1.5 to S7.5. Control (CTL) denotes the experiment with $A_c$=0. In the aquaplanet configuration, the N and corresponding S experiments are mirror images of each other. Each S–N pair thus differs only in the direction of the periodic forcing relative to the mean ITCZ, so that the difference between the two experiments directly measures the directionality of the pole-to-pole teleconnection.

The Qflux is imposed for 220 years, 22 forcing cycles. Although $Q_p$ neither adds nor extracts energy over the long-term average, nonlinear cloud feedbacks could instigate a weak adjustment of the global-mean surface temperature (Shaw *et al* 2015, Shin *et al* 2017). The first two forcing cycles are hence discarded as spin-up. We then apply a Lanczos low-pass filter with a 5-year cutoff period to the remaining 20 cycles and construct the composite periodic response. Unless stated otherwise, the amplitude of a periodic response denotes half the peak-to-peak range of the composite cycle, and polar quantities are cosine-weighted averages over 70–90° of the respective hemisphere.

## 3. Results

The zonal-mean hovmöller diagram displays the canonical pattern of the pole-to-pole teleconnection in the CTL experiment (Fig. 1d-g). The Qflux imposed at the northern-latitude drives a decadal oscillation in the NH polar SST (>70°N; $T_N$) with an amplitude of 4.7 K. The warming propagates equatorward and displaces the ITCZ in phase with the forcing; the ITCZ, defined as the precipitation centroid between 20°S and 20°N (Donohoe *et al* 2013), fluctuates between 4.18°S and 4.22°N (cyan and green lines in Fig. 1d,e). This propagation takes about 4 months, measured as the lag of maximum correlation between the SST at the $Q_p$ boundary (i.e., 40°N) and at the climatological ITCZ, consistent with previous studies once scaled by the mixed-layer depth (Woelfle *et al* 2015). The equatorward propagation is strongly limited right at the ITCZ where the Hadley cell return flow converges, often called the blocking effect (Kang *et al* 2020). Although the direct SST response is largely confined to the forced hemisphere, a weak but distinct decadal response emerges at the opposite pole through atmospheric circulation adjustments (Fig. 1g). As the ascending branch of the Hadley cell is shifted northward, the southern Hadley cell strengthens (purple contours in Fig. 1g) and transports more angular momentum into the subtropics, accelerating the subtropical jet. The stronger subtropical jet reinforces the vertical wind shear on its poleward flank, and the enhanced baroclinicity pulls the eddy-driven jet equatorward (Lee and Kim 2003, Brayshaw *et al* 2008). The resulting eddy momentum divergence induces anomalous descent, which adiabatically warms the polar troposphere and, in turn, the surface (Fig. 1f).

We then investigate how the background climatology modulates the pole-to-pole teleconnection. Figures 2a-c are shown for three representative climates in which the mean ITCZ is in the unforced hemisphere (S6.0), near the equator (CTL), and in the forced hemisphere (N6.0). The forced-hemisphere response is nearly identical in the three experiments (Fig. 2a-c): the $T_N$ propagates equatorward, is blocked at the climatological ITCZ of each experiment, and displaces the ITCZ in phase with the forcing. This behavior extends across the series of experiments. Despite the shift in the mean ITCZ, the $T_N$ amplitude remains within a narrow range of 4.7 to 5.1 K, a spread of less than 9% (Fig. 2e). Only S7.5 falls outside this range (5.5 K), likely because its colder polar background favors stronger polar amplification (e.g., Kim *et al* 2018). The ITCZ likewise oscillates about its own climatological position with a comparable amplitude of 3.5 to 4.4°, regardless of whether the ITCZ is defined by the precipitation centroid or by the Hadley cell ascending branch (Figs. 2g and S1).

No matter where the climatological ITCZ lies, a periodic SST oscillation emerges at the opposite pole (Fig. 2a–c). Its amplitude, however, differs. As the mean ITCZ shifts southward, the SST amplitude in the southern high-latitude ($T_S$) increases systematically from 0.29 to 0.61 K (Fig. 2d; left axis of Fig. 2f). This dependence is not explained by the ITCZ amplitude, which varies marginally across the experiments; a smaller ITCZ amplitude rather accompanies a larger $T_S$ (Fig. 2g). The forced-polar response itself varies slightly with the background state (Fig. 2e), so we quantify the teleconnection by pole-to-pole efficiency as the regression coefficient of $T_S$ onto $T_N$. Note that we use the lag of maximum correlation diagnosed in CTL and apply it consistently to all experiments. The efficiency follows the absolute $T_S$ response, increasing from 0.045 to 0.111 K $K^{-1}$ (right axes of Fig. 2f,g). The pole-to-pole teleconnection is therefore directionally asymmetric: it is about twice as efficient when the mean ITCZ lies in the unforced hemisphere (S7.5 versus N7.5).

What process, then, limits the teleconnection initiated in the hemisphere of the mean ITCZ (i.e., the summer-like hemisphere)? To identify this, we decompose the pole-to-pole efficiency into three sensitivities representing the sequential steps of the teleconnection: the ITCZ displacement per unit $T_N$ anomaly (L1), changes in SH subtropical jet strength ($STJ_{SH}$) per unit ITCZ displacement (L2), and the $T_S$ response to per unit $STJ_{SH}$ anomaly (L3). Each term is a lagged regression coefficient, as is the efficiency. Unlike the ratio of amplitudes, the product of the three regression coefficients does not necessarily reproduce the overall efficiency. If it reproduces the efficiency well, it suggests the sequential pathway is predominant (Runge *et al* 2015). As shown in Fig. 3a, the product closely reproduces the efficiency across the experiments so the directional asymmetry can be attributed to the individual sensitivities.

We find the linkage between the ITCZ and STJSH as the primary source of directional asymmetry in the pole-to-pole teleconnection. The L1, which measures the ITCZ response to high-latitude SST changes, is modulated by the background climatology. However, it peaks when the mean ITCZ lies near the equator and decreases on either side and thereby does not account for the directional asymmetry (Fig. 3b). This nonmonotonic dependence differs from the monotonic one reported by Collavini *et al* (2026), in which the ITCZ response strengthens as the climatological ITCZ approaches the extratropical cooling owing to cloud radiative effects. The difference may arise from their one-directional ITCZ shift under steady forcing, from the cloud response of the model, or from both. The L3 encompasses the eddy-driven jet shift, the accompanying secondary circulation, and subsequent polar feedbacks. The climatological jet location could have modulated this link, as a jet closer to the equator tends to shift more under global warming (e.g., Kidston and Gerber 2010), but L3 shows no systematic dependence on the mean ITCZ within the range of climates examined here (Fig. 3d). Given a comparable $STJ_{SH}$ response, pole-to-pole teleconnection thereby emerges with similar magnitude regardless of the background climatology. In contrast, L2 decreases systematically as the climatological ITCZ shifts northward (Fig. 3c): a comparable ITCZ displacement, accompanied by a comparable change in the southern Hadley cell strength (Fig. S2), drives a progressively weaker $STJ_{SH}$ response. This dependence holds for both phases of the forcing, the acceleration with a northward shift and the deceleration with a southward shift, with no sign of saturation in either phase (Fig. S3). Thus, these results suggest that the directional asymmetry arises primarily from how efficiently changes in the tropical circulation accelerate or decelerate the subtropical jet.

This efficiency can be understood from angular momentum conservation in the upper branch of the Hadley cell. Air ascending at the ITCZ carries the angular momentum of that latitude poleward, so that the STJ response to an ITCZ shift has two contributions: how closely the Hadley cell conserves angular momentum and how much angular momentum is supplied at the ITCZ, which is largest at the equator. For a northward ITCZ shift, the former strengthens the $STJ_{SH}$ regardless of the ITCZ location, whereas the latter depends it. It adds angular momentum when the ITCZ moves toward the equator and vice versa. To quantify the two contributions, we adapt the angular-momentum scaling of Kang and Lu (2012). The scaling theory begins with the local Rossby number (Ro), defined as the negative ratio of relative vorticity $\zeta$ to planetary vorticity $f$ (Walker and Schneider 2006),

$$Ro \equiv -\zeta/f\,, \quad \zeta = -\frac{1}{a\cos\varphi}\left(\frac{\partial}{\partial\varphi}u\cos\varphi\right), \quad f = 2\Omega\sin\varphi, \tag{3}$$

where $\varphi$ is latitude, $u$ is the zonal-mean zonal wind, and $\Omega$ and $a$ are the rotation rate and radius of Earth. Ro is a nondimensional measure of the proximity of the tropical circulation to the inviscid, angular-momentum conserving regime. The cross-equatorial winter cell has larger Ro as angular momentum is effectively conserved in its upper branch, whereas the summer cell has smaller Ro as it is more strongly influenced by eddy momentum fluxes (Walker and Schneider 2006). Integrating Eq. (3) from the ITCZ latitude $\varphi_i$ to the southern Hadley cell edge $\varphi_h$ gives

$$u_h\cos\varphi_h - u_i\cos\varphi_i = 2\Omega a\int_{\varphi_i}^{\varphi_h} Ro(\varphi)\sin\varphi\cos\varphi\, d\varphi, \tag{4}$$

where $u_i$ and $u_h$ are the zonal winds at the ITCZ and SH Hadley cell edge, respectively. We approximate the SH subtropical jet strength by $u_h$. Similar to Kang and Lu (2012), we also neglect $u_i$ and adopt a single effective Rossby number for southern Hadley cell $R_{SH}$. Then, Eq. (4) becomes

$$u_h \approx \frac{\Omega a}{\cos\varphi_h} R_{SH}(\varphi_i)(\sin^2\varphi_h - \sin^2\varphi_i). \tag{5}$$

As the mean ITCZ shifts into the NH, the southern Hadley cell becomes cross-equatorial and approaches the angular-momentum-conserving regime, motivating the representation of its effective dynamical regime as a function of ITCZ latitude, $R_{SH}(\varphi_i)$. Differentiating with respect to $\varphi_i$ at fixed $\varphi_h$ provides a theoretical estimate of the $STJ_{SH}$ sensitivity to the ITCZ shift, that is, the scaled L2:

$$\frac{du_h}{d\varphi_i} \approx \underbrace{\frac{\Omega a}{\cos\varphi_h} R'_{SH}(\sin^2\varphi_h - \sin^2\varphi_i)}_{\text{regime effect}} \underbrace{- \frac{\Omega a}{\cos\varphi_h} R_{SH}\sin(2\varphi_i)}_{\text{source effect}}. \tag{6}$$

The first term, the regime effect, is the $STJ_{SH}$ change through the change in the Hadley cell regime. The second term, the source effect, represents the $STJ_{SH}$ change through the shift of the upper branch drawing angular momentum; it changes sign at the equator. We diagnose $R_{SH}(\varphi_i)$ from 27 additional experiments with various extratropical forcings (Supplementary Text S1; Fig. S4). We first average the local Rossby number within the upper branch of the southern Hadley cell and away from the equator, where $f$ is small: 300 hPa over 22°–12°S in every experiment. This band is chosen a posteriori for which the scaled and diagnosed ITCZ–STJ sensitivities agree best (Fig. S4e). We then fit a quadratic function to the averaged Rossby number as a function of mean ITCZ latitude to obtain $R_{SH}(\varphi_i)$. Note that fitting $R_{SH}(\varphi_i)$ instead of the CTL experiment alone and applying it to all eleven climatologies gives consistent scaled sensitivity (Fig. S5). The southern Hadley cell edge $\varphi_h$ is fixed at its mean diagnosed latitude of approximately 27°S, as it varies little across the experiments.

The scaled sensitivity captures the decrease in L2 as the mean ITCZ shifts northward (Fig. 4a). Integrating this sensitivity with respect to ITCZ latitude consistently reproduces the nonlinear ITCZ-STJ relationship in the periodic experiments, with steeper slopes of S experiments and flatter slopes of N experiments (Fig. 4b). The source effect accounts for most of the contrast in the scaled sensitivity, whereas the regime effect remains positive and varies relatively little across the mean ITCZ positions (inset of Fig. 4a). Thus, although the regime effect contributes to the overall subtropical jet response, the directional asymmetry arises primarily because an ITCZ displacement moves the angular momentum source toward the equator in one climatology and away from it in the other (Fig. 5).

## 4. Summary and discussion

In this study, we examine how the zonal-mean atmospheric pole-to-pole teleconnection depends on the background climatology using an aquaplanet model coupled to a slab ocean. We impose a steady interhemispheric forcing to displace the mean ITCZ and a periodic forcing over the northern high-latitude. The teleconnection occurs in every climatology but is about twice as efficient when the mean ITCZ is in the unforced hemisphere; it is hence directionally asymmetric with respect to the ITCZ. Decomposing the efficiency identifies the ITCZ–STJ link as the step where the asymmetry arises, and an angular-momentum scaling explains why. The subtropical jet is fed by the angular momentum that air carries at the ITCZ, and this supply is largest when the ITCZ is at the equator. Even for a comparable Hadley cell response, the jet response to an ITCZ shift depends on the distance of the ITCZ from the equator. The ITCZ sets to reduce the interhemispheric energy asymmetry, but the geographic equator does not, so the same northward shift moves the ITCZ closer to the equator in one climatology and farther from it in another. The same shift therefore reinforces the jet response in one climatology and partially offsets it in the other, resulting the directional asymmetry.

While the idealized configuration is key to identifying the directional asymmetry, the same simplifications entail caveats that need to be considered when interpreting the results, particularly with realistic configuration. For example, L3 shows no systematic dependence on the mean ITCZ in our climatologies, but with topography, land–sea contrast, ocean circulation, or sea ice, it is likely to differ between the Arctic and Antarctic, amplifying or damping the directional asymmetry. In addition, the zonal-mean pathway operates together with the Rossby wave teleconnection and ocean adjustment, so that how these processes modulate the directional asymmetry remains to be examined. We also note that the scaling is diagnostic rather than predictive. The effective Rossby number for southern Hadley cell is fitted a posteriori, and although we allow a quadratic dependence on the ITCZ latitude, the fitted dependence is nearly linear within the ±4° range examined here. Although this range covers the intermodel spread (Byrne *et al* 2018), the Hadley cell regime may respond differently for a larger ITCZ displacement or in a realistic configuration, in which case the regime effect may also contribute to the directional asymmetry.

These findings provide a dynamical basis for interpreting the fast pole-to-pole teleconnection recorded in ice cores. During the last glacial period, the ITCZ was displaced southward relative to today (Schneider *et al* 2014). In our framework, such a background state itself favors the north-to-south teleconnection—the direction in which the decadal-scale linkage is actually observed: the abrupt Greenland changes shifted the southern westerlies and the Antarctic temperature within decades (Markle *et al* 2017, Buizert *et al* 2018). The background state may also contribute to the stronger Antarctic response to abrupt Greenland warming than to cooling (Svensson *et al* 2026) although this asymmetry is not directly examined in our experiments. While this asymmetry partly reflects differences in the magnitude and abruptness of the Greenland forcing, with stadial-to-interstadial warming generally being larger and more abrupt than interstadial-to-stadial cooling, the ITCZ was also farther south during stadials than during interstadials, at least in the tropical Atlantic (e.g., Peterson et al. 2000; Deplazes et al. 2013), so warming transitions, which start from stadials, begin from a background state that favors the north-to-south teleconnection.

Since the present-day ITCZ lies north of the equator, our results imply that south-to-north teleconnections may be more efficient than north-to-south teleconnections, although this remains to be tested in a realistic climate configuration. This tendency is qualitatively consistent with recent evidence that Antarctic-to-Arctic atmospheric teleconnections can be more effective than Arctic-to-Antarctic teleconnections (Moon *et al* 2026), although the asymmetry there is not attributable to the zonal-mean pathway alone but also involves tropical Pacific coupling. The zonal-mean ITCZ is projected to shift only slightly under global warming (Byrne *et al* 2018). Once the warming is mitigated, however, the ITCZ is projected to shift southward, even cross the equator, because of the delayed response of the AMOC and the Southern Ocean (Kug *et al* 2022, Oh *et al* 2024, Wang *et al* 2026). Our results suggest that such a transition would alter the directional efficiency of the teleconnection. This matters particularly for large high-latitude perturbations expected in a warming world, such as ice-sheet collapse and the accompanying meltwater discharge (e.g., Winkelmann *et al* 2026), and for regional climate interventions, such as stratospheric aerosol injection (e.g., Jeon *et al* 2026), whose remote impacts would then depend on the background climatology at the time. Whether this dependence emerges in the real climate warrants further study, but our results identify the climatological ITCZ as a key factor regulating the directional efficiency of the atmospheric pole-to-pole teleconnection.

**Acknowledgments**

We thank Jaeheung Park and Hanjun Kim for their helpful discussions during the early stages of this work. This study is supported by the National Research Foundation of Korea (NRF) grant funded by the Korea government (MSIT) (NRF-2022R1A3B1077622; RS-2024-00334637).

**Conflict of interest**

The authors declare no competing interests.

**Data Availability Statement**

The dataset supporting the conclusions of this article will be available in the Zenodo repository.

**Author contributions**

Y.S. conceived the study, designed and analyzed the experiments, and wrote the manuscript. J.-S.K. supervised the study and edited the manuscript.

**Code Availability Statement**

All codes required to reproduce the study will be available in the Zenodo repository.

## Figure list

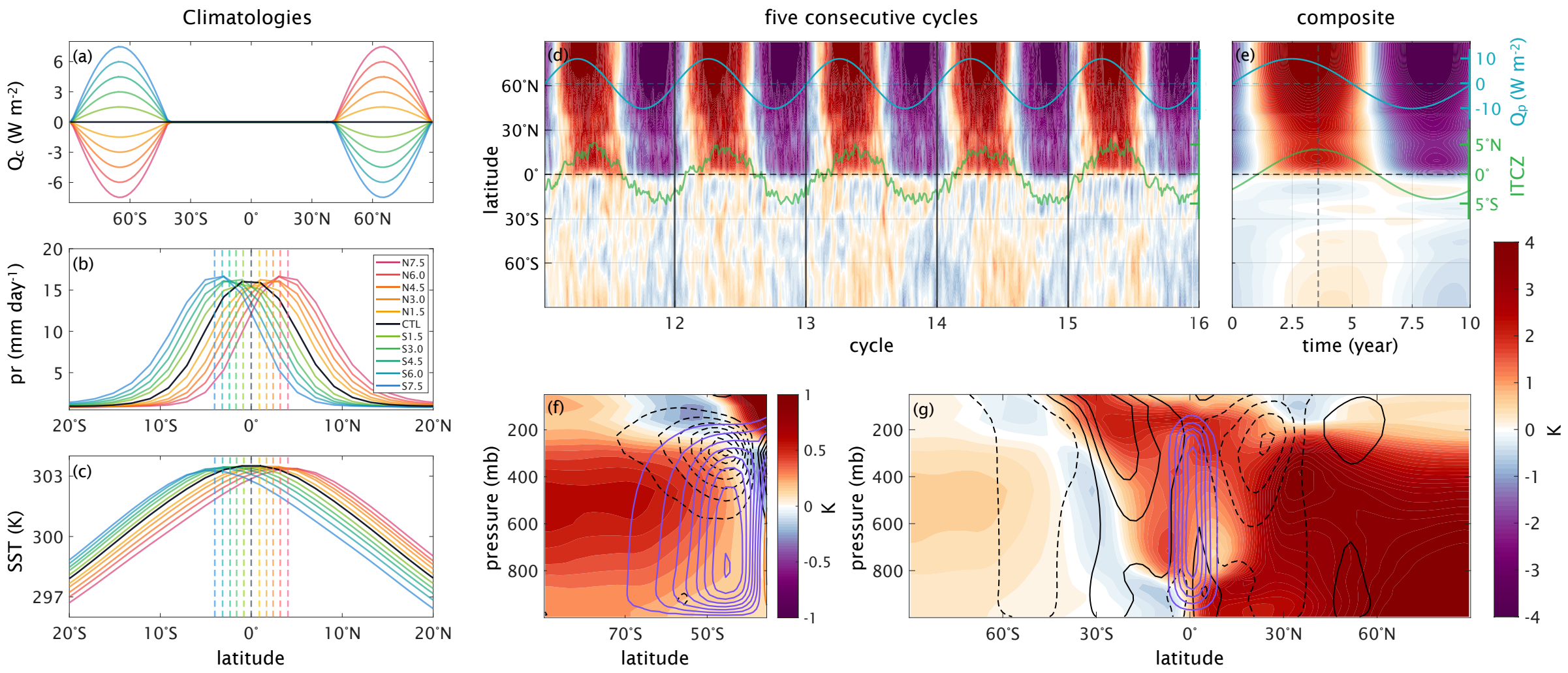

**Figure 1**. Background climatologies and the pole-to-pole teleconnection in CTL. (a) Hemispherically asymmetric Qflux $Q_c$ (W m$^{-2}$), (b) climatological zonal-mean precipitation (mm day$^{-1}$), and (c) SST (K) in the 11 experiments. Dashed vertical lines denote the mean ITCZ latitude of each experiment. (d) Hovmöller diagram of the zonal-mean SST response (shading; K) in CTL for five consecutive cycles (i.e., 50 years) and (e) composite across $Q_p$ cycles. $Q_p$ is overlaid as a cyan line (right axis; W m$^{-2}$) and the ITCZ latitude as a green line (right axis; °). The ITCZ is defined as the precipitation centroid between 20°S and 20°N. (f, g) Vertical cross-sections of the temperature response (shading; K) at maximum northward ITCZ shift (dashed line in e). (f) For the southern extratropics, meridional streamfunction (clockwise in orange and counterclockwise in purple; interval = 0.8 × 10$^9$ kg s$^{-1}$) and transient eddy momentum flux convergence (solid-dashed contour; interval = 0.1 m s$^{-1}$ day$^{-1}$) are overlaid. (g) For the globe, meridional streamfunction (clockwise in orange and counterclockwise in purple; interval = 0.5 × 10$^{11}$ kg s$^{-1}$) and zonal-mean zonal wind (solid-dashed contour; interval = 2 m s$^{-1}$) are overlaid. Note the different color scales in (f) and (g).

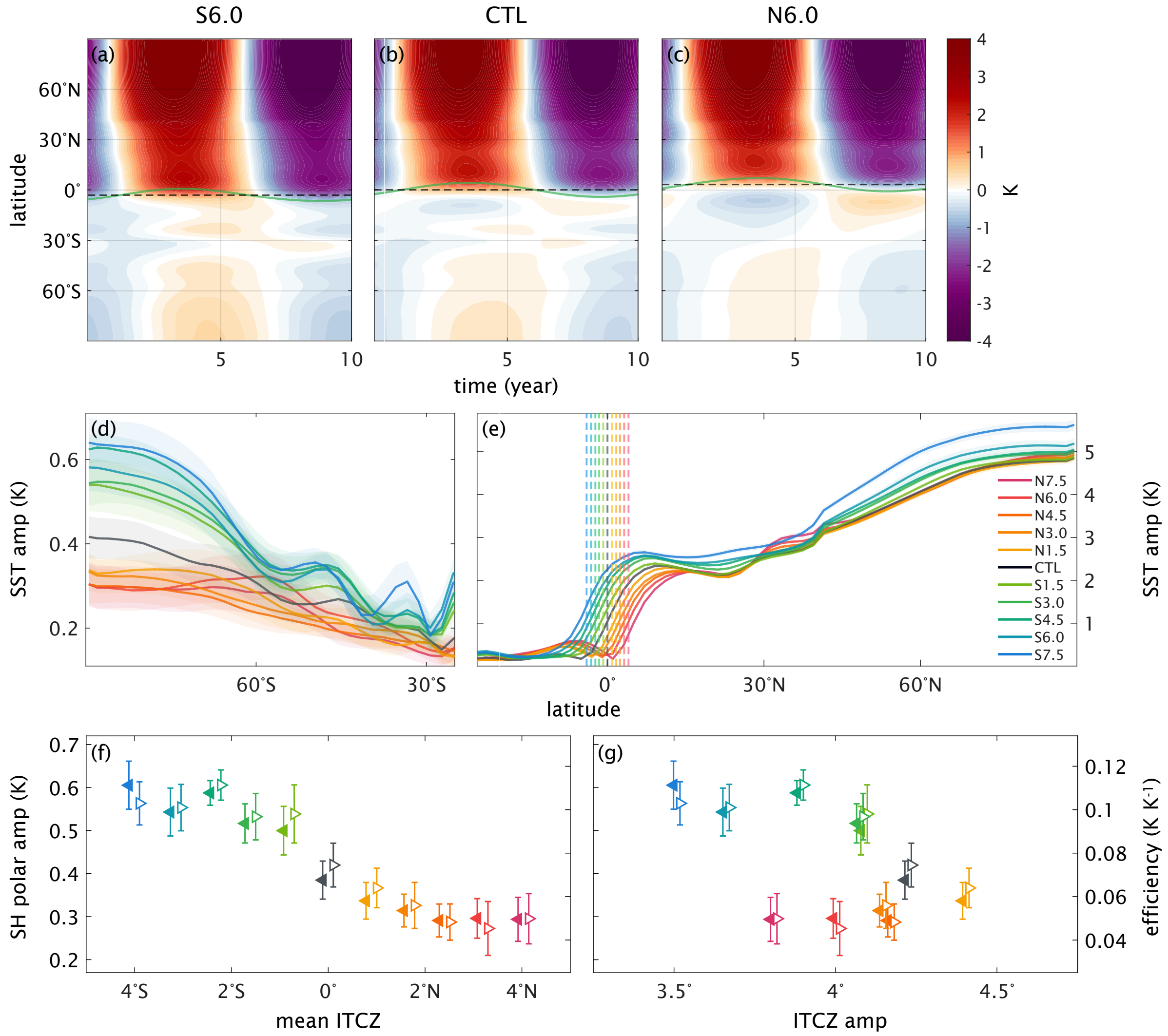

**Figure 2**. Directional asymmetry of the pole-to-pole teleconnection. (a–c) Composite Hovmöller diagrams of the zonal-mean SST response (shading; K) and ITCZ latitude (green line; °) in S6.0, CTL, and N6.0. The horizontal dashed line denotes the climatological ITCZ of each experiment. (d, e) Amplitude of the SST response (K) in the (d) Southern (unforced) and (e) Northern (forced) Hemisphere. Shading denotes the 95% confidence level across the 20 forcing cycles. Dashed vertical lines denote the climatological ITCZ latitude. (f, g) Amplitude of $T_S$ (filled triangles; left axis) and the pole-to-pole efficiency (open triangles; right axis) as a function of (f) the mean ITCZ and (g) the ITCZ amplitude. Error bars denote the 95% confidence level based on a Student's t-test across the forcing cycles.

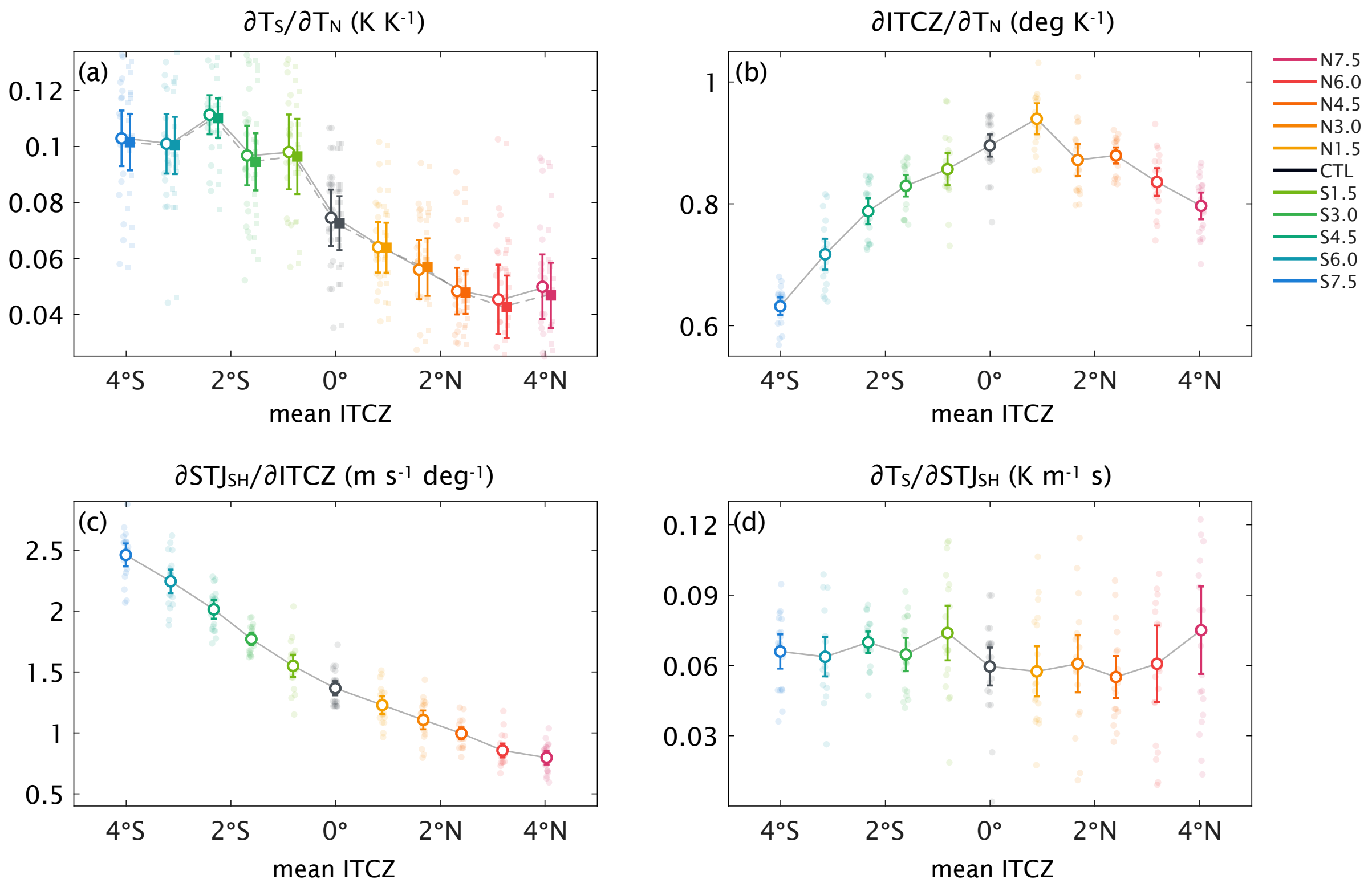

**Figure 3**. Decomposition of the pole-to-pole efficiency into three sensitivities. (a) Pole-to-pole efficiency $\partial T_S/\partial T_N$ (open circles) and the product of the three sensitivities L1 × L2 × L3 (filled squares) as a function of the mean ITCZ. (b–d) The three sensitivities: (b) L1, ITCZ shift per unit $T_N$; (c) L2, $STJ_{SH}$ response per unit ITCZ shift; and (d) L3, $T_S$ response per unit $STJ_{SH}$ response. Each sensitivity is the lagged regression coefficient computed in each forcing cycle, with the lag of maximum correlation diagnosed in CTL (4, 0, and 10 months for L1, L2, and L3, and 12 months for the efficiency) and applied to all experiments. Faded dots denote individual cycles, open circles the mean, and error bars the 95% confidence level based on a Student's t-test.

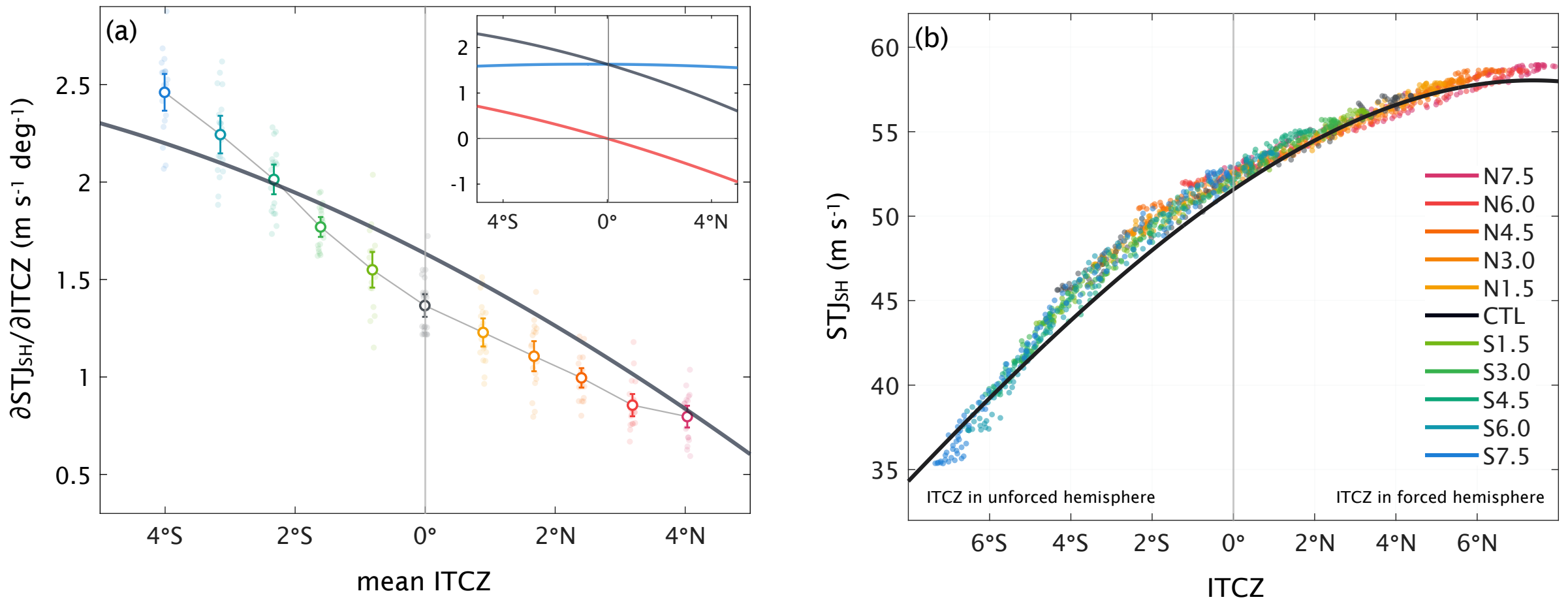

**Figure 4**. Momentum scaling of the ITCZ–$STJ_{SH}$ sensitivity. (a) As in Figure 3c, but with the scaled sensitivity from Eq. (6) overlaid (black line). The inset decomposes the scaled sensitivity (black) into the regime effect (blue) and the source effect (red). (b) Composite $STJ_{SH}$ against ITCZ latitude in all experiments (dots) and the scaled $STJ_{SH}$ obtained by integrating Eq. (6) from the CTL mean state (thick black line).

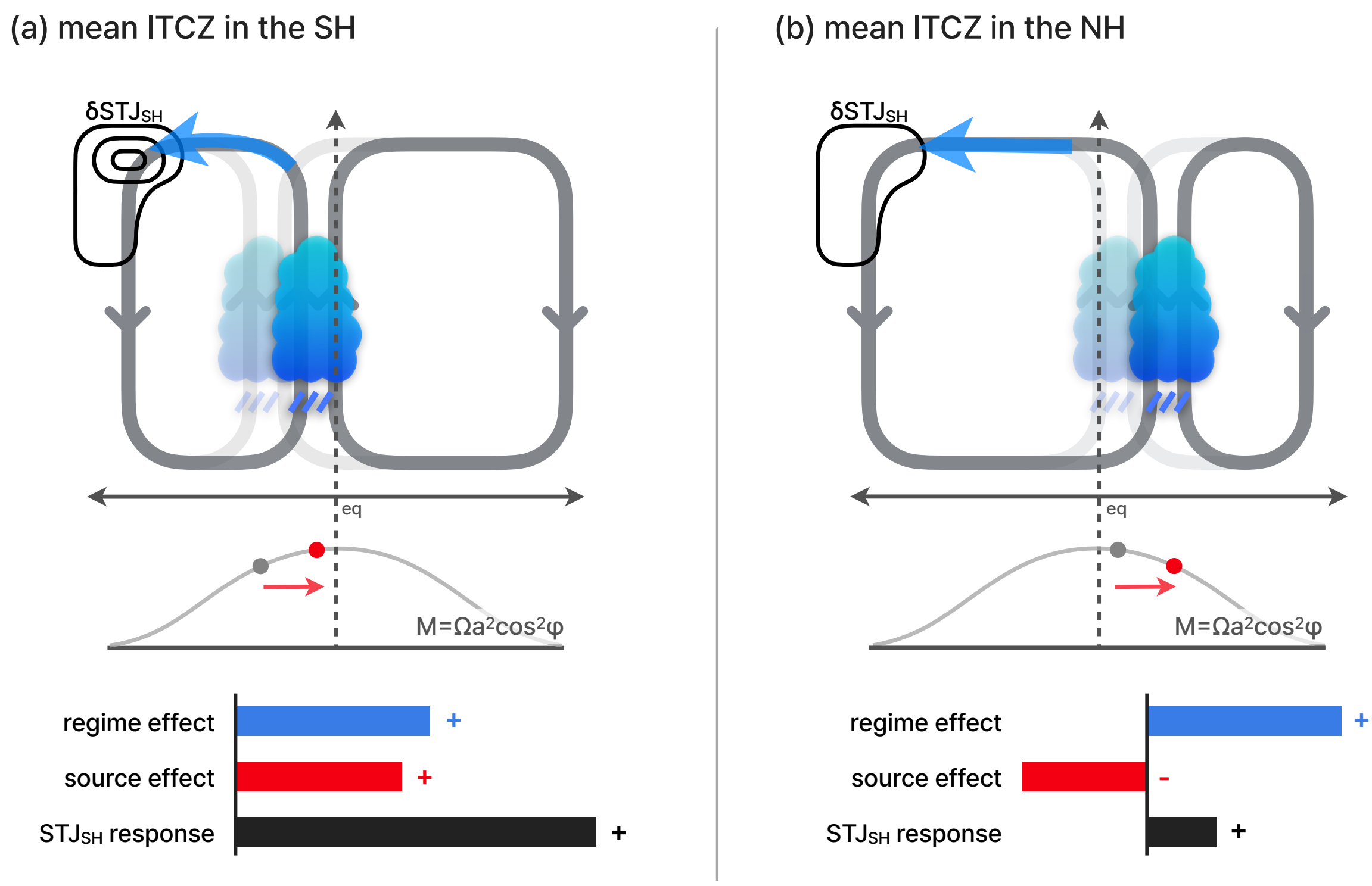

**Figure 5**. Schematic of the directional asymmetry: the southern subtropical jet (STJ_SH) response to the same northward ITCZ shift when the mean ITCZ lies in (a) the SH and (b) the NH. Light and dark gray loops denote the Hadley cells before and after the shift. The blue arrow denotes the enhanced poleward transport of angular momentum by the strengthened southern Hadley cell, which occurs in both cases (regime effect), and the black contours the resulting $STJ_{SH}$ acceleration. The planetary angular momentum, $M = \Omega a^2 cos^2\varphi$, is shown as a gray line with the ITCZ before (gray) and after (red) the shift. The shift moves the ITCZ toward the equator in (a) and away from it in (b), increasing or decreasing the angular momentum supplied to the upper branch (source effect). The bar chart summarizes the regime effect (blue), the source effect (red) and their sum, the STJ_SH response, which reinforce each other in (a) and partially offset in (b).